\documentclass[%
 reprint,
 superscriptaddress,
 amsmath,amssymb,
 prl,
floatfix,
]{revtex4-2}

\usepackage{graphicx}
\usepackage{dcolumn}
\usepackage{bm}
\usepackage{array}
\usepackage[hidelinks]{hyperref}
\usepackage{booktabs}
\usepackage{upgreek}
\usepackage{siunitx}

\usepackage[
separate-uncertainty = true,
multi-part-units=single
]{siunitx}
\usepackage{xcolor}
\usepackage[version=4]{mhchem}
\usepackage[normalem]{ulem}
\usepackage{tablefootnote}

\usepackage[section]{placeins} 

\usepackage{lineno}

\newcommand{\RevA}[1]{{\color{black}#1}}

\newcommand\sonec{S1${c}$}
\newcommand\stwoc{S2${c}$}

\DeclareSIUnit\keVnr{keV_{nr}}
\DeclareSIUnit\keVee{keV_{ee}}

\newcommand\XeOneTwoNine{\ce{^{129}Xe}}

\newcommand\XeOneThreeOne{\ce{^{131}Xe}}

\newcommand\supplementalurl{https://www.hepdata.net/record/155182}

\begin{document}
\title{Dark Matter Search Results from 4.2 Tonne-Years of Exposure \\ of the LUX-ZEPLIN (LZ) Experiment} 
\author{J.~Aalbers}
\affiliation{SLAC National Accelerator Laboratory, Menlo Park, CA 94025-7015, USA}
\affiliation{Kavli Institute for Particle Astrophysics and Cosmology, Stanford University, Stanford, CA  94305-4085, USA}

\author{D.S.~Akerib}
\affiliation{SLAC National Accelerator Laboratory, Menlo Park, CA 94025-7015, USA}
\affiliation{Kavli Institute for Particle Astrophysics and Cosmology, Stanford University, Stanford, CA  94305-4085, USA}

\author{A.K.~Al Musalhi}
\affiliation{University College London (UCL), Department of Physics and Astronomy, London WC1E 6BT, UK}

\author{F.~Alder}
\affiliation{University College London (UCL), Department of Physics and Astronomy, London WC1E 6BT, UK}

\author{C.S.~Amarasinghe}
\affiliation{University of California, Santa Barbara, Department of Physics, Santa Barbara, CA 93106-9530, USA}

\author{A.~Ames}
\affiliation{SLAC National Accelerator Laboratory, Menlo Park, CA 94025-7015, USA}
\affiliation{Kavli Institute for Particle Astrophysics and Cosmology, Stanford University, Stanford, CA  94305-4085, USA}

\author{T.J.~Anderson}
\affiliation{SLAC National Accelerator Laboratory, Menlo Park, CA 94025-7015, USA}
\affiliation{Kavli Institute for Particle Astrophysics and Cosmology, Stanford University, Stanford, CA  94305-4085, USA}

\author{N.~Angelides}
\affiliation{Imperial College London, Physics Department, Blackett Laboratory, London SW7 2AZ, UK}

\author{H.M.~Ara\'{u}jo}
\affiliation{Imperial College London, Physics Department, Blackett Laboratory, London SW7 2AZ, UK}

\author{J.E.~Armstrong}
\affiliation{University of Maryland, Department of Physics, College Park, MD 20742-4111, USA}

% ...

\collaboration{The LZ Collaboration} 

\begin{abstract}
We report results of a search for nuclear recoils induced by weakly interacting massive particle (WIMP) dark matter using the LUX-ZEPLIN (LZ) two-phase xenon time projection chamber. This analysis uses a total exposure of $4.2\pm0.1$~tonne-years from 280~live days of LZ operation, of which $3.3\pm0.1$~tonne-years and 220 live days are new. A technique to actively tag background electronic recoils from $^{214}$Pb $\beta$ decays is featured for the first time. Enhanced electron-ion recombination is observed in two-neutrino double electron capture decays of $^{124}$Xe, representing a noteworthy new background. After removal of artificial signal-like events injected into the data set to mitigate analyzer bias, we find no evidence for an excess over expected backgrounds. World-leading constraints are placed on spin-independent (SI) and spin-dependent WIMP-nucleon cross sections for masses $\geq$9~GeV/$c^2$. The strongest SI exclusion set is $2.2\times10^{-48}$~cm$^{2}$ at the 90\% confidence level and the best SI median sensitivity achieved is $5.1\times10^{-48}$~cm$^{2}$, both for a mass of 40~GeV/$c^2$.
\end{abstract}

\maketitle

Astrophysical observations provide strong evidence for the existence of dark matter and its dominance of the matter density of the Universe~\cite{Aghanim2020, SofueRubin, harvey2015nongravitational, ARBEY2021103865}. Well-motivated as a thermal relic, the weakly interacting massive particle (WIMP) is a leading hypothesis for dark matter, arising in a number of beyond-the-standard-model theories~\cite{RevModPhys.90.045002, Billard2022,akerib2022snowmass2021}. The LUX-ZEPLIN (LZ) collaboration operates the world’s largest two-phase xenon time projection chamber (TPC), with the primary aim of directly detecting WIMPs through their interactions with xenon nuclei. This technology has produced the most stringent constraints on spin-independent (SI) WIMP-nucleon interactions for $\gtrsim$5~GeV/$c^2$ masses, with all three current-generation experiments having reported results~\cite{SR1WS, XENON:2023cxc, PandaX:2024qfu}. In this Letter we present the latest WIMP analysis from LZ, combining a new 220~live-day exposure (WS2024) with the 60~live-day exposure from the first result (WS2022)~\cite{SR1WS} to perform the most sensitive direct search for WIMPs to date.

The LZ experiment~\cite{lzcdr,*lztdr,*LZExperiment} is located 4850~ft underground, shielded by a 4300~meter water equivalent rock overburden in the Davis Cavern at the Sanford Underground Research Facility (SURF) in Lead, South Dakota, USA. Its TPC, containing seven tonnes of active liquid xenon (LXe), is surrounded by a dual-detector veto system comprising a two-tonne LXe ``Skin," designed to tag $\gamma$ rays, and an outer detector (OD) containing 17 tonnes of gadolinium-loaded liquid scintillator (GdLS), optimized for the detection of neutrons. 

Interactions in the LXe target that fills the TPC create light in the form of prompt vacuum ultraviolet (VUV) scintillation (S1) and delayed electroluminescence (S2), both observed by the top and bottom photomultiplier tube (PMT) arrays. The S2 signal is generated by ionization electrons as they traverse a thin layer of gaseous xenon above the liquid, having been drifted upwards from the interaction site under an applied electric field and extracted from the liquid phase. The measured time interval between the S1 and the S2 pulses, referred to as the ``drift time”, corresponds to the depth, $z$, of the interaction, while the hit pattern of S2 light in the top PMT array allows for position reconstruction in the transverse $(x,y)$ plane. The ratio of S2 to S1 signal sizes is used to discriminate between background electronic recoils (ERs) and nuclear recoils (NRs), the latter of which is expected for WIMP-nucleus scatters. Background events are further rejected if they exhibit multiple scatters (MS) rather than a single scatter (SS) in the TPC, or through the presence of coincident signals in the Skin and/or OD.

The LXe target is contained in a vacuum-insulated cryostat, and is continuously circulated for gas-phase purification to remove electronegative impurities, and to regulate the liquid level. The cooling and circulation systems allow for control of the LXe flow pattern inside the cryostat, which influences the spatial distribution and movement of radioisotopes in the TPC volume. Two distinct flow configurations can be attained, referred to as the ``high-mixing" and ``low-mixing" states. In the high-mixing state, the liquid flow results in spatially homogeneous distributions of injected calibration sources. In the low-mixing state, there exist distinct regions of laminar flow throughout the detector, and coincident $^{222}$Rn$-^{218}$Po $\alpha$ decays can be reliably used to map the LXe velocity field. The ability to model the flow in this state is exploited in the development of a ``radon tag" to target the prominent $^{214}$Pb ER background (similar to that in Ref.~\cite{XENON_radonVeto}), discussed later.

Cathode, gate, and anode electrodes establish the TPC drift field at 97~V/cm and the liquid-gas extraction field at 3.4~kV/cm (at the radial center, below the liquid surface) for the WS2024 data-taking campaign. With respect to WS2022, the extraction field was decreased to reduce the dead time incurred from spurious electron and photon emission, and the cathode voltage was lowered in response to the onset of persistent light emission localized to the Skin region below the cathode.

This analysis incorporates 220.0~live days collected in the new electrode bias configuration between March~27,~2023 and April~1,~2024. Of this WS2024 live time, 40.9~live days are included from when the detector is in the high-mixing state, due to proximity to injected calibrations, with the remaining 179.1~live days taken in the low-mixing state. The xenon gas pressure throughout the run is 1.860 bar(a) with variation at the level of 0.03\%. The level of convective mixing is controlled by adjusting the temperature of the liquid returning to the bottom of the TPC from 177.9~K (high-mixing state) to 174.2~K (low-mixing state), with variations at the level of 0.02\% for each. Changing from high-mixing to low-mixing state increases the maximum drift time from 1045~$\upmu$s to 1050~$\upmu$s and reduces the overall liquid height by about 0.2 mm. The consequent 1\% increase in S2 signal size is corrected for in analysis. \RevA{Stable electron lifetimes
averaging 9.3 ms and 13.8 ms were maintained in the
high-mixing and low-mixing states, respectively.}

Data acquisition uses a trigger with a digital filter designed primarily to capture S2 pulses. The trigger was optimized for WS2024 to increase the sensitivity to low-energy events of interest to the WIMP analysis, achieving an efficiency of 95\% for S2 pulses corresponding to 3.5~extracted electrons. Waveforms from the TPC, Skin, and OD PMTs are recorded with a window extending from 2.2~ms before the trigger time to 2.5~ms after it, sampled at 100~MHz. Further details of the data acquisition and signal chain are described in Ref.~\cite{LZ_DAQ}.

Event waveforms are processed to find and categorize pulses and interactions. Pulse sizes are normalized by the time-dependent size of single VUV photon pulses measured in each channel, and are reported as the number of photons detected (phd)~\cite{faham2015measurements, LOPEZPAREDESdpe}. Events are categorized as SS if they have a single S1 preceding a single S2. The SS identification efficiency is dependent on the S1 and S2 pulse classification efficiencies. Ionization electrons diffuse as they drift, which can lead to fragmentation of the S2 signal into multiple pulses. The splitting probability is $<$1\% on average across the TPC for S2 pulses relevant to this analysis, as estimated with waveform simulations~\cite{LZSims}. For a pulse to be classified as S1, coincident photons must be recorded in at least three PMTs; however, undersized PMT responses and differences in photon transit times can lead to missed S1 pulses. The SS S1 identification efficiency is found to be $>$90\% for all S1 pulse sizes in this analysis, rising to 100\% for S1s $>$35~phd (8~keV). This efficiency is assessed by comparing the event classification of a sample of tritium beta decays and deuterium-deuterium (DD) neutron calibration events determined by the event reconstruction algorithms to the corresponding classification obtained from a manual, visual categorization campaign.

Tritium and DD neutrons are among several sources used to calibrate the TPC~\cite{LZ_calibrations}. Tritium is delivered through radiolabeled methane along with $^{14}$C, in an activity ratio of approximately 8:1. The tritiated methane, primarily used to calibrate the ER response, was injected into the TPC while in the high-mixing state to ensure its efficient dispersal and complete removal \RevA{by the purification system, with a half-life of 53 hours}. Mono-energetic 2.45~MeV neutrons from a collimated DD generator, directed roughly 10~cm below the LXe surface, supply the counterpart for calibrating the NR response. AmLi neutron sources, deployed in source tubes at three different azimuthal positions and three different heights around the TPC, provide additional NR calibration events. Injections of $^{131{\rm m}}$Xe and $^{83{\rm m}}$Kr at roughly monthly and trimonthly cadences, respectively, in addition to background radon progeny, provide mono-energetic peaks throughout WS2024, used to monitor the stability of the detector response and develop corrections for the normalization of S1 and S2 signals.

S1 and S2 signals exhibit spatial and temporal variations due to factors such as changing detector conditions or non-uniform light collection and electron extraction efficiencies. Corrections are applied to produce standardized values, \sonec{} and \stwoc{}. An $(x,y,z)$ map normalizes the S1 signals to the center of the TPC volume, with correction factors averaging 8\% in absolute value. S2 signals are normalized to the radial center of the liquid surface in the $(x,y)$ plane, in addition to being corrected for the \RevA{measured electron lifetime}; the average values of these corrections are approximately 13\% and 7\%, respectively. The responses are also normalized in time, using the tritium calibration period as the reference. 

Calibrations also inform the tuning of the \textsc{nest} 2.4.0~\cite{nest_2_4_0} model for detector response, in which the scintillation photon and ionization electron gains, $g_1$ and $g_2$, are key parameters. A scan over viable values for $g_1$ and $g_2$, performed using tritium data in $\{{\rm S1}c, \text{log}_{10}({\rm S2}c) \}$ space and constrained by the energy reconstruction of $^{131{\rm m}}$Xe and $^{83{\rm m}}$Kr peaks, yields $g_1 = 0.112\pm0.002$~phd/photon and $g_2 = 34.0\pm0.9$~phd/electron. A further tuning to more accurately capture the ``ER band," i.e. the shape of the observed distribution of tritium data, enables \textsc{nest} to reproduce the band median and width to better than 0.2\% and 5\% in $\log_{10}({\rm S2}c)$, respectively. The DD data define an analogous ``NR band," for which \textsc{nest} is separately tuned to a similar level of precision via small adjustments in the NR yield parameters. \RevA{For both the ER and NR models, the tuned parameters and the resulting light and charge yields are within the uncertainty bands reported in Ref.~\cite{szydagis2022review}.} The DD neutron interactions are confined to the top of the detector, where the average electric field is 7\% higher than the center. Because the charge yield of $^{222}$Rn $\alpha$ decays used for spatial corrections is more field-sensitive than NRs, corrections in the DD neutron region overestimate field effects on recombination, leading to a systematic shift in the NR band position as assumed for the full TPC. An upwards shift of 2.4\% in \stwoc{} is applied to the NR model to compensate for this effect, calculated from electrostatic simulations and \textsc{nest}-based models.

Candidate events for the WIMP analysis are selected if they are classified as SS; they are located within the fiducial volume (FV); they pass veto coincidence, S1 and S2-based, and live-time exclusion cuts, all of which are described below; and they lie within the WIMP region of interest (ROI). The ROI is defined such that the corrected S1 and S2 pulse sizes are in the range of $3<{\rm S1}c<80$~phd and $\log_{10}({\rm S2}c)<4.5$. A lower bound on uncorrected S2 pulse size of 645~phd, equivalent to 14.5 extracted electrons, is also applied. This S2 threshold renders the analysis largely insensitive to $^8$B coherent elastic neutrino-nucleus scattering (CE$\nu$NS), a search for which will be reported in a future publication.

\begin{figure}[tbp]
 \centering
 \includegraphics[ width=\linewidth]{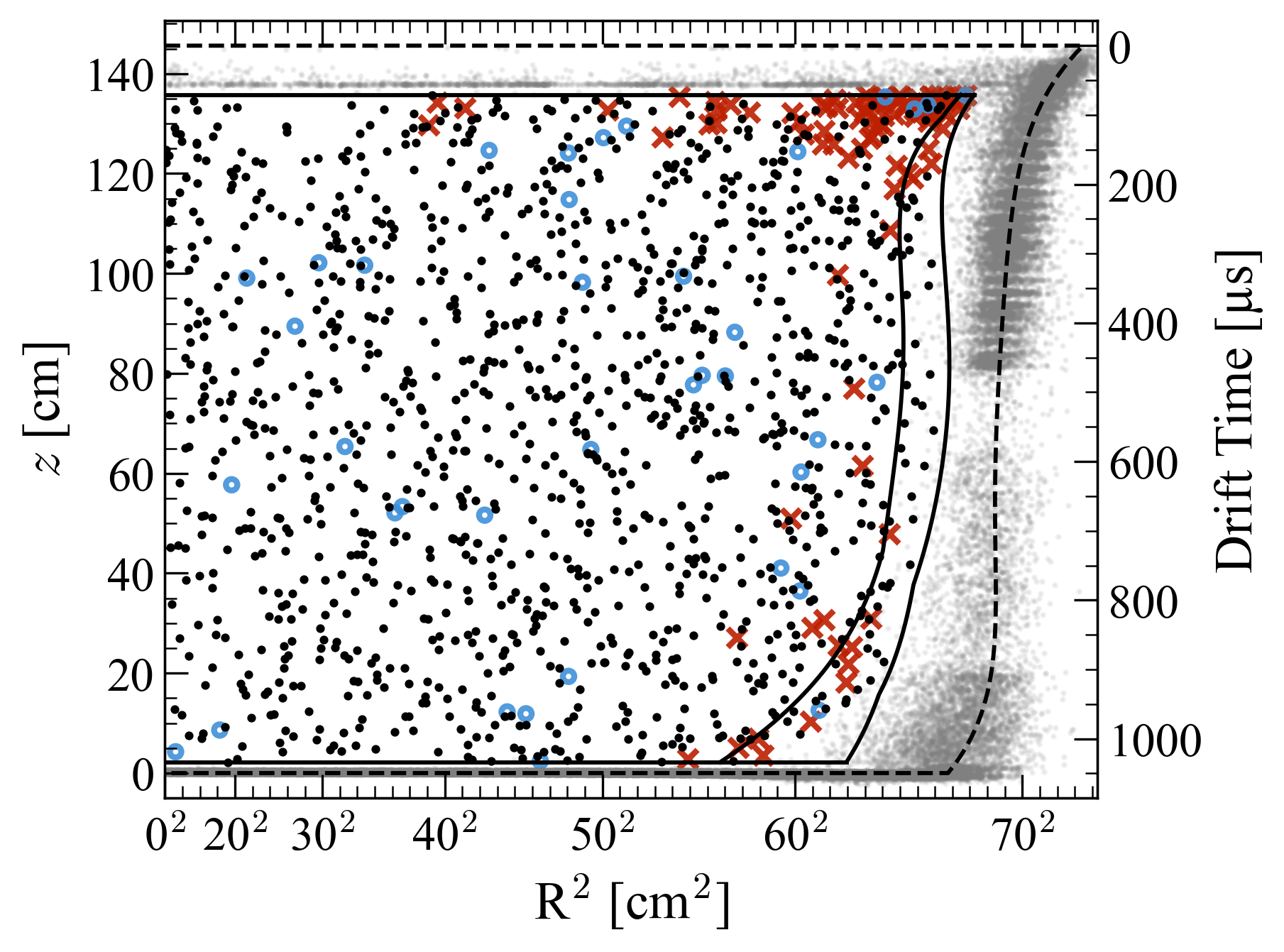}
 \caption{Data from WS2024, in black with all analysis cuts applied, and in gray outside the FV without veto coincidence cuts. Red crosses and blue circles represent events that are vetoed by the prompt and delayed veto coincidence cuts, respectively. The dashed line shows the active volume, averaged over azimuth, and the two solid lines depict the FV, at the azimuths of its smallest and largest radial extent. The drift field affects the position reconstruction such that the observed radius of the TPC varies as a function of both drift time and azimuthal angle.}
 \label{fig:r2z}
\end{figure}

A FV, illustrated in Fig.~\ref{fig:r2z}, is defined to suppress external backgrounds. Drift time limits are placed at 71~$\upmu$s and 1034 (1030)~$\upmu$s in the low(high)-mixing state, corresponding to distances of 9.9~cm below the gate and 2.2~cm above the cathode, respectively. Events within 6.0~cm of the TPC field cage resistors~\cite{LZExperiment} in $(x,y)$ are also excluded, as these resistors have elevated radioactivity. The radial fiducial boundary is defined such that backgrounds originating from the TPC wall, which can be misreconstructed radially inwards due to finite position resolution, have a negligible expectation of $<0.01$ events for WS2024. This radial boundary follows a contour of measured wall background event rate, and is a function of both depth and azimuthal angle. The fiducial mass, determined using the fraction of tritium events in the full active volume found within the FV, is $5.5\pm0.2$~tonnes.

Veto coincidence cuts reject events that are accompanied by activity in the Skin and/or OD, and are divided into prompt and delayed categories (see Fig.~\ref{fig:r2z}). Prompt cuts target $\gamma$ rays and fast neutron scatters, removing events with signals in the Skin (OD) of size $>2.5$~phd ($>4.5$~phd) within 0.25~$\upmu$s (0.3~$\upmu$s) of the S1. Delayed veto cuts target neutrons, which thermalize then capture, typically on Gd in the OD with the release of an 8--9~MeV $\gamma$-ray cascade. These cuts reject events with pulses in the OD (Skin) of energy $\gtrsim200$~keV ($\gtrsim300$~keV) that occur within a 600~$\upmu$s time window after the S1. The delayed veto cuts remove 3\% of the collected live time due to coincidences between unrelated interactions in the TPC and the veto detectors. The veto efficiency for AmLi calibration neutrons, combining prompt and delayed cuts, is measured to be $89\pm3$\%. Radiogenic neutrons can be vetoed more effectively as they are typically higher in energy than those from AmLi, and are often accompanied by other neutrons and $\gamma$ rays. Simulations tuned on the AmLi calibration data indicate that the efficiency to reject background neutrons is $92\pm4$\%, with the uncertainty driven by residual discrepancies between the simulations and AmLi data.

Cuts based on S1 and S2 pulse characteristics target accidental coincidence events, in which the random pairing of pulses classified as S1s and S2s mimic true SS interactions. Several sources give rise to isolated S1- and S2-classified pulses, such as interactions in light- or charge-insensitive regions of the detector, spurious Cherenkov photon emissions, or delayed emissions of electrons~\cite{sorensen2017electron,LUXelectronbkgs}. The cuts primarily exploit the shapes of these pulses in assessing the validity of the S1-S2 pair. The rejection power of the cuts is evaluated on a sideband of known accidental coincidence events with unphysical drift times ($>$1100~$\upmu$s) and is 99.5\% for all such events in the WIMP ROI.

Live time exclusions discount periods with higher-than-average incidences of accidental coincidences due to elevated pulse rates, which generally occur in the wake of high-energy interactions. A time hold-off is imposed after large S2 signals, dependent on their size, to avoid subsequent delayed ionization or correlated light emission. The hold-off calculation is optimized with respect to that used in WS2022 so that it cumulatively removes 10\% of the recorded live time in WS2024, as opposed to 30\% in WS2022. Accounting for all live-time losses and exclusions, 81.5\% of the collected live time of WS2024 remains for analysis, totaling 220.0~live days.

\begin{figure}[t]
\center
\includegraphics[width=\columnwidth]{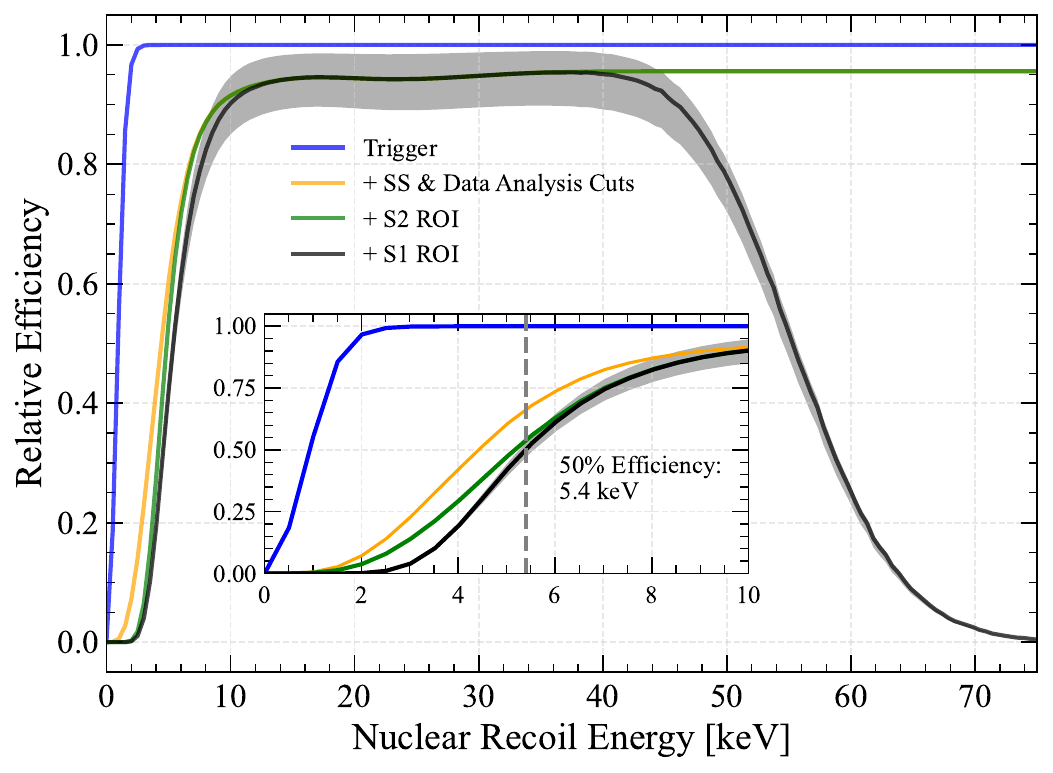}
\caption{Energy-dependent NR signal efficiency in WS2024 after the application, in sequence, of the S2 trigger (blue); SS reconstruction and analysis cuts (orange); S2 ROI (green) followed by the S1 ROI (black). The inset shows the low-energy behavior, with the dotted line at 5.4~keV marking 50\% efficiency. The uncertainty band (gray) is assessed using AmLi and tritium data as discussed in the text.
}
\label{fig:NRacceptance}
\end{figure}

The cumulative effect of the data selection, event detection, and reconstruction efficiencies on the NR signal efficiency can be seen in Fig.~\ref{fig:NRacceptance}, where the cut acceptances are assessed on a combination of tritium and neutron calibration data sets. Uncertainties arise from the combined statistical error on the SS efficiency and data analysis cuts, as well as a systematic error equal to the variation in efficiency of data analysis cuts as evaluated on the different calibration data sets.

To mitigate analyzer bias, artificial, WIMP-like events are injected at random times into the data pipeline, a process referred to as ``salting." Salt events are pre-generated by combining S1 and S2 pulses from sequestered tritium and AmLi calibration data sets. The pulse sizes for a given event are chosen to best match those expected from an NR energy deposition drawn from a parent distribution consisting of the sum of an exponential (WIMP-like) and a flat spectrum. The salt injection rate and parameters of the parent energy distribution were generated randomly within predefined bounds and kept hidden for the entirety of the analysis. Eight salt events were injected, of which one was removed by analysis cuts, consistent with the NR signal efficiency. The other seven events were revealed as salt following the final definition of the data selection criteria and the likelihood models. \RevA{More details about the salting distribution can be found in the Supplemental Material~\cite{Supplemental}.}

\begin{figure}[t]
\center
\includegraphics[width=\columnwidth]{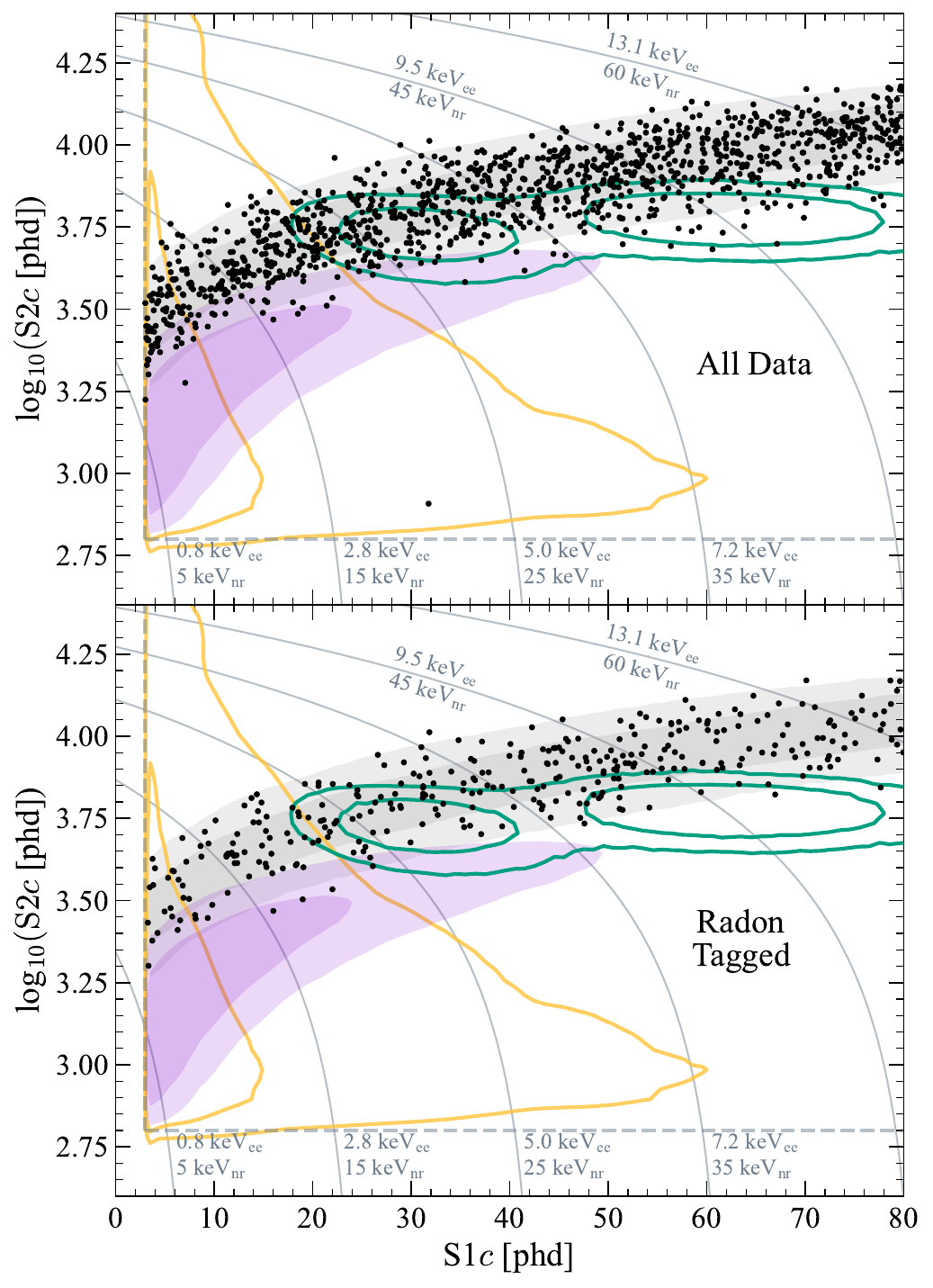}
\caption{Black points show the final set of events passing all selection cuts with the panels distinguishing all events in WS2024 (top) from those that are radon tagged (bottom). The radon-tagged data set contains 60\% and 15\% of the full FV activity of the $^{214}$Pb and dispersed backgrounds, respectively; thus, the comparison with all data brings to prominence the non-$^{214}$Pb backgrounds like $^{124}$Xe that are not well-described by the ER band. Dark and light gray and purple shading show the regions containing 68\% and 95\% of the ER portion of the background model and a 40~GeV/$c^2$ WIMP, respectively. Green and orange contours show the same quantiles for the best-fit $^{124}$Xe distribution and the accidental coincidence events, respectively. Gray lines show contours of constant ER-equivalent (keV$_{\rm ee}$) and NR-equivalent (keV$_{\rm nr}$) energy.
}
\label{fig:data}
\end{figure}

The 1220 events which remain in the 3.3~tonne-year exposure of WS2024 after the application of all data selections and salt removal are shown in Figs.~\ref{fig:r2z} and~\ref{fig:data}. These events are analyzed using a likelihood model in the $\{{\rm S1}c, \text{log}_{10}({\rm S2}c) \}$ observables. Signal and background model distributions are generated using \textsc{nest} and event simulations~\cite{LZSims}, with the exception of the accidental coincidence background model, which follows a data-driven method identical to that used for the WS2022 result~\cite{LZ-backgrounds}. The primary backgrounds and their treatment are similar to those of the WS2022 analysis, with the expected counts for each background source given in Table~\ref{tab:backgrounds}. Notable differences are now described. 

Two sources of background NR events are considered: those originating from neutrino interactions and those from radioactive decays in detector components. The neutrino interactions are from CE$\nu$NS of solar ($^{8}$B and \textit{hep}) and cosmic-ray-generated atmospheric neutrinos. Nuclear recoils that originate from spontaneous fission and $(\alpha, n)$ reactions in detector materials can be vetoed by detecting the scattered neutrons and/or associated $\gamma$ rays in the Skin and/or OD. Events vetoed by these systems are used as a sideband to derive an \textit{in-situ} constraint on the number of neutron events in the WIMP search data. As discussed later, the sideband is incorporated directly into the final statistical analysis; a sideband-only fit returns an expectation of $0^{+0.2}$ SS counts, as shown in Table~\ref{tab:backgrounds}. Secondary estimates from simulations based on detector material radioassays and from data using candidate MS neutron events tagged by the veto detectors return expectations of $0.05 \pm 0.01$ and $0.3\pm0.2$ neutron counts, respectively, consistent with the primary analysis.

Electron capture (EC) and double electron capture (DEC) decays deposit their energy via X ray and Auger cascades in a more localized profile than a $\beta$ particle of the equivalent energy, leading to higher ionization densities and enhanced electron-ion recombination. As a result, EC and DEC events look more NR-like than $\beta$ decays, as reported in Ref.~\cite{Xelda-Lshell}. Preliminary measurements of this effect for ECs in LZ were performed with sidebands of Skin-tagged and multiple scatter decays of $^{127}$Xe and $^{125}$Xe, isotopes that have been produced \textit{in situ} via neutron activation. These yield a charge-yield ratio of $Q_y^\text{L} / Q_y^\beta=0.87\pm0.03$ for 5.2~keV L-shell EC decays relative to $\beta$ decays of the same energy~\cite{aalbers2025measurementsmodelsenhancedrecombination}.

Two-neutrino DEC decays of $^{124}$Xe are observed in LZ with a half-life of ${T_{1/2} = (1.09 \pm 0.14_{\text{stat}} \pm 0.05_{\text{sys}}) \times \qty{e22}{yr}}$~\cite{LZ-Xe124}, consistent with that in Ref.~\cite{XENONnT-Xe124_Xe136-2022}. The expected number of $^{124}$Xe counts in the ROI is estimated assuming an abundance of $9.52\times10^{-4}$~\cite{XeAbundances} and branching ratios for the LL (10.00~keV) and LM (5.98~keV) decay modes of 1.4\% and 0.8\%, respectively. The LM charge suppression is modeled with the same ratio of 0.87 as used for single EC decays. The overlap of two L-shell cascades in LL-shell capture should result in additional recombination. As an independent measurement of the LL-shell charge yield is not available, the Thomas-Imel (TI) box model ~\cite{thomas1987recombination, szydagis2022review} is invoked to relate the ionization density to the amount of electron-ion recombination at a given applied electric field. The ratio of the charge yield for LL-shell DEC to $\beta$ decays of the same energy, $Q_y^\text{LL} / Q_y^\beta$, is treated as a free parameter in the statistical analysis, bounded on one side by the measured $Q_y^\text{L} / Q_y^\beta$ for WS2024 (0.87) and on the other side by the result from a preliminary study of increasing the ionization density in the TI box model (0.65).

The largest ER background contribution in this exposure comes from $\beta$-emitting isotopes in the LXe bulk. Xenon sampling measurements indicate a concentration of $1.10\pm0.18$~ppb $^\text{nat}$Ar g/g and an exposure-averaged concentration of $186\pm26$~ppq $^\text{nat}$Kr g/g. These are used to estimate the contributions of $^{39}$Ar and $^{85}$Kr, respectively. Exposure in the high-mixing state includes residual tritium and $^{14}$C activity following \RevA{their injection for ER} calibrations, and both their beta spectra are therefore incorporated into the background model for this period. 

Radon emanates from materials exposed to xenon, dispersing in the active volume. With a measured FV activity of $3.9\pm0.4$~$\upmu$Bq/kg, $^{214}$Pb decays from the $^{222}$Rn chain, in particular those to the ground state, form the dominant ER background. \RevA{We attribute the higher activity in WS2024 with respect to WS2022 ($3.26~\upmu$Bg/kg \cite{LZ-backgrounds}) to the lower drift field, as charged radon daughter progeny move more slowly out of the fiducial volume.} A method to identify events with high likelihood of being a $^{214}$Pb decay --- the radon tag --- is developed using simulations of the movement of daughter atoms and ions built from observed flow vectors between $^{222}$Rn-$^{218}$Po $\alpha$-decay pairs. A tag volume in which it is probable to find a $^{214}$Pb decay is defined for each identified $^{218}$Po alpha decay, co-moving with the projected flow streamline for 81~min or approximately three times the $^{214}$Pb half-life. \RevA{The tag volume includes two flowpaths to account for neutral and ionized daughters. The radon tag can only be applied in the low-mixing state, where the flow can be modeled. Additional periods in the low-mixing state during which radon tagging is not possible, for example around gaps in data acquisition or after flow disturbances induced by power outages, are labeled as radon-tag-inactive. 
To summarize, we have four categories of WS2024 exposure: (1) high-mixing, where a radon tag is not possible; (2) low-mixing, but with the radon tag inactive due to interruptions; (3) low-mixing, active radon tag and radon-tagged, with a high probability to contain $^{214}$Pb decays; and (4) low-mixing, active radon tag and radon-untagged, with a lower probability to contain $^{214}$Pb decays.} 
 
 \RevA{For WS2024, 15\% of the low-mixing, tag-active exposure is identified as radon-tagged, containing on average $60\pm4$\% of the total $^{214}$Pb as estimated directly by measurements of its excited-state decays.}
The effective $^{214}$Pb activity in the untagged remainder of the tag-active exposure is thus $1.8\pm0.3$~$\upmu$Bq/kg. \RevA{The radon-tagged data set can be seen in the bottom panel of Fig.~\ref{fig:data}.}
The separation into \RevA{radon-tagged and untagged} exposures offers the benefit of being able to better identify and constrain other background contributions, $^{124}$Xe in particular. 

\begin{table}[hbtp]
    \caption{The expected and best-fit counts for different sources in the 3.3~tonne-year exposure of WS2024, including a 40~GeV/$c^2$ WIMP signal. Flat $\beta$-decay components have been separated based on whether they are affected by the flow state or the radon tag, and the neutron counts are derived \textit{in situ}. ``Det. $\gamma$s" refers to $\gamma$-ray contributions from detector materials, whose Compton spectra are also flat in the ROI.}
    \label{tab:backgrounds}
    \centering
    \begin{tabular}{lr@{}lr@{}l}
    \tabularnewline
    \hline
    \hline
    Source & \multicolumn{2}{l}{Pre-fit Expectation}
    & \multicolumn{2}{c}{Fit Result}\tabularnewline
    \hline \tabularnewline[-2.2ex]
    \ce{^{214}Pb} $\beta$s &\centering 743 &~$\pm$ 88\phantom{0} &\centering  733 &~$\pm$ 34 \phantom{0}  
    \tabularnewline
    $^{85}$Kr + $^{39}$Ar $\beta$s + det. $\gamma$s &\centering 162  &~$\pm$ 22\phantom{0} &\centering 161 &~$\pm$ 21 \phantom{0} 
    \tabularnewline
    Solar $\nu$ ER& 102 &~$\pm$ 6 \phantom{0} & 102 &~$\pm$ 6 \phantom{0}
    \tabularnewline
    $^{212}$Pb + $^{218}$Po $\beta$s &\centering 62.7 &~$\pm$ 7.5 \phantom{0} &\centering 63.7 &~$\pm$ 7.4 \phantom{0} 
    \tabularnewline
    Tritium+ $^{14}$C $\beta$s &\centering 58.3  &~$\pm$ 3.3 \phantom{0} &\centering 59.7 &~$\pm$ 3.3 \phantom{0}
    \tabularnewline
    \ce{^{136}Xe} $2\nu\beta\beta$ &	55.6 &~$\pm$ 8.3 \phantom{0} & \RevA{55.9} &~$\pm$ 8.2 \phantom{0} 
    \tabularnewline 
    \ce{^{124}Xe} DEC & 19.4 &~$\pm$ \RevA{2.5}  & \RevA{20.4} &~$\pm$ \RevA{2.4} 
    \tabularnewline
    \ce{^{127}Xe} +  \ce{^{125}Xe} EC & 3.2 &~$\pm$ 0.6  & 2.7 &~$\pm$ 0.6 
    \tabularnewline
    Accidental coincidences	& 2.8 &~$\pm$ 0.6 & 2.6 &~$\pm$ 0.6 
     \tabularnewline
    ${\rm Atm.}$ $\nu$ NR & 0.12 &~$\pm$ 0.02 & 0.12 &~$\pm$ 0.02
    \tabularnewline
    $^8$B$+hep$ $\nu$ NR & 0.06 &~$\pm$ 0.01 & 0.06 &~$\pm$ 0.01  
    \tabularnewline
   \hline \tabularnewline[-2.2ex]
    Detector neutrons &	\multicolumn{2}{l}{\phantom{00}\footnotemark[1]$0.0^{+0.2}$} &  \multicolumn{2}{c}{$0.0^{+0.2}$} \tabularnewline[0.25ex]
    40~GeV/$c^2$ WIMP &\hfill &\phantom{0}--&  \multicolumn{2}{c}{$0.0^{+0.6}$} 
    \tabularnewline[0.25ex]
    \hline \tabularnewline[-2.2ex]
    Total & 1210 &~$\pm$ 91 & \RevA{1202} &~$\pm$ \RevA{41} \phantom{0} \tabularnewline
    \hline
    \hline
    \end{tabular}\\
      \footnotetext[1]{The expected number of neutron events results from a fit to the sample of veto detector-tagged events. This expectation is not explicitly used in the final combined fit as this sample is included directly in the likelihood, as described in the text.
      }
\end{table}

The data are divided into six mutually exclusive samples, which are simultaneously analyzed using an unbinned, extended maximum likelihood. Five of these apply to the new 220~live-day exposure and describe: \RevA{the (1-4) WS2024 samples described previously}
and (5) events tagged by the Skin or OD. The nuisance parameters, which float within their uncertainties, are the expected number of events for each background source, listed in Table~\ref{tab:backgrounds}; the radon tagging efficiency; the neutron veto tagging efficiency; and the 
$^{124}$Xe $Q_y^\text{LL} / Q_y^\beta$, bounded between 0.65 and 0.87. Parameters characterizing $^{222}$Rn and $^{220}$Rn daughter backgrounds, tritium and $^{14}$C are independent in the high- and low-mixing state exposures. The sixth sample in the likelihood represents WS2022, containing the data and models from the first, 60~live-day WIMP search result~\cite{SR1WS}. \RevA{The full WS2024 data set is shown in the top panel of Fig.~\ref{fig:data}. All six samples can be seen in Fig.~S1 of the Supplemental Material~\cite{Supplemental}, where we also report their exposures.}

\begin{figure}[t]
\center
\includegraphics[width=\columnwidth]{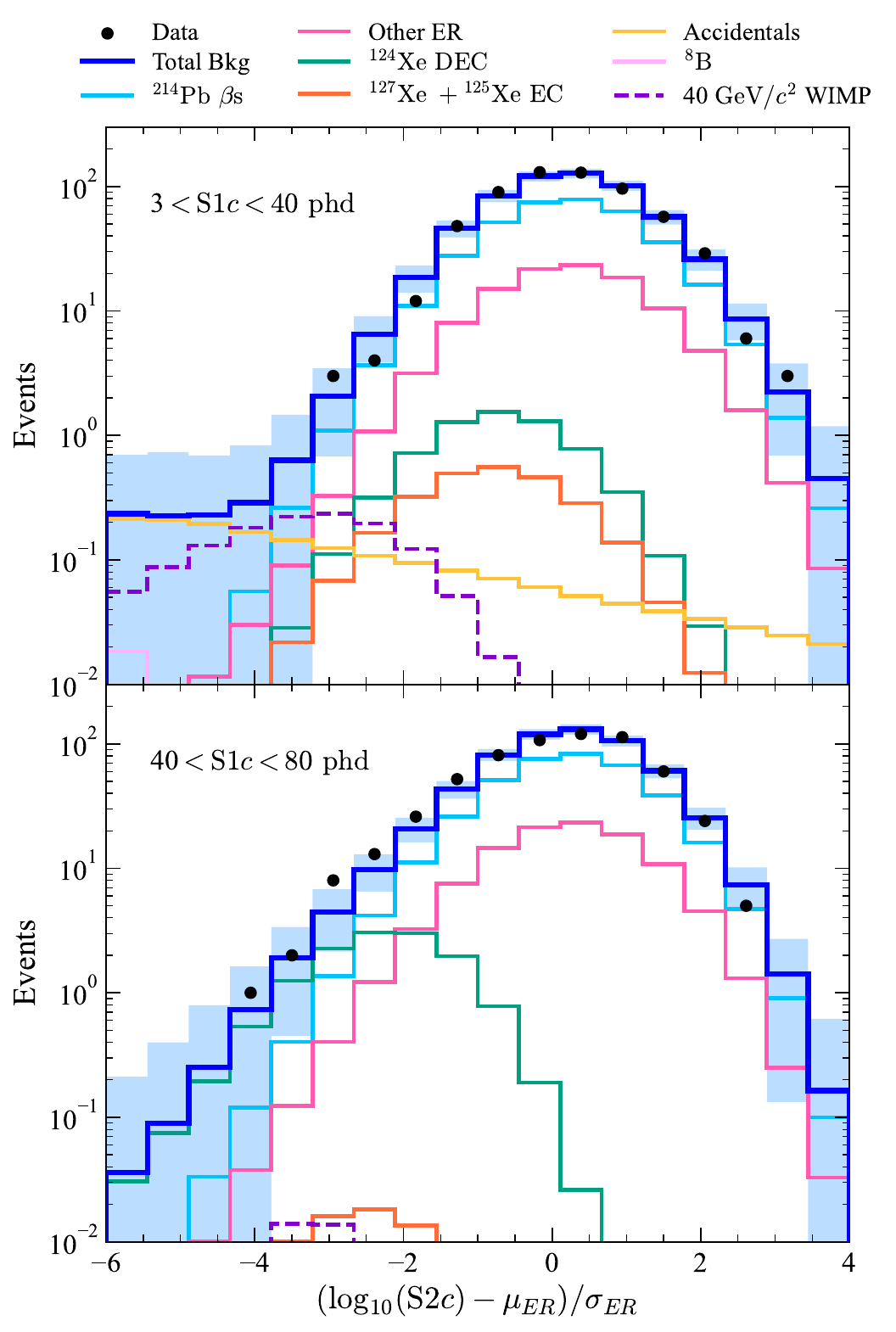}
\caption{Comparison between data and the best-fit model for WS2024 corresponding to the values shown in Table~\ref{tab:backgrounds} for a background plus a 40~GeV/$c^2$ WIMP signal fit in a projection along the ER band median, normalized to the $1\sigma$ band width. The 40~GeV/$c^2$ WIMP component is shown with an amplitude equal to its power-constrained 90\% C.L. upper limit. The upper and lower panels select the \sonec{} ranges 3--40~phd and 40--80~phd with $p$ values of 0.55 and \RevA{0.59}, respectively. Shaded bands depict the central interval containing 68\% of the combined systematic and statistical uncertainties of the model.}
\label{fig:gof}
\end{figure}

The best-fit number of WIMPs at all masses tested (between 9~GeV/$c^2$ and 100~TeV/$c^2$) for the combined WS2024+WS2022 analysis is zero. The goodness of fit of the background-only model is assessed in binned 1D projections of the data and model into ${\rm S1}c$, $\text{log}_{10}({\rm S2}c)$, reconstructed energy, distance from the ER band median, as well as the 2D space of $\{{\rm S1}c, \text{log}_{10}({\rm S2}c) \}$. All samples show excellent model-data agreement, passing a Holm-Bonferroni test~\cite{Holm1979ASS} with a significance level of 0.05. Figure~\ref{fig:gof} shows the WS2024 data and model in the 1D projection along the ER band median, illustrating the level of agreement through the tails of the distribution with the inclusion of the modified $^{124}$Xe response. The fitted values of the radon tagging efficiency and the $^{124}$Xe $Q_y^\text{LL} / Q_y^\beta$ are consistent across all masses, at $62\pm3$\% and $0.70\pm0.04$, respectively, in agreement with the pre-fit expectations.

\begin{figure}[t]
\center
\includegraphics[width=\columnwidth]{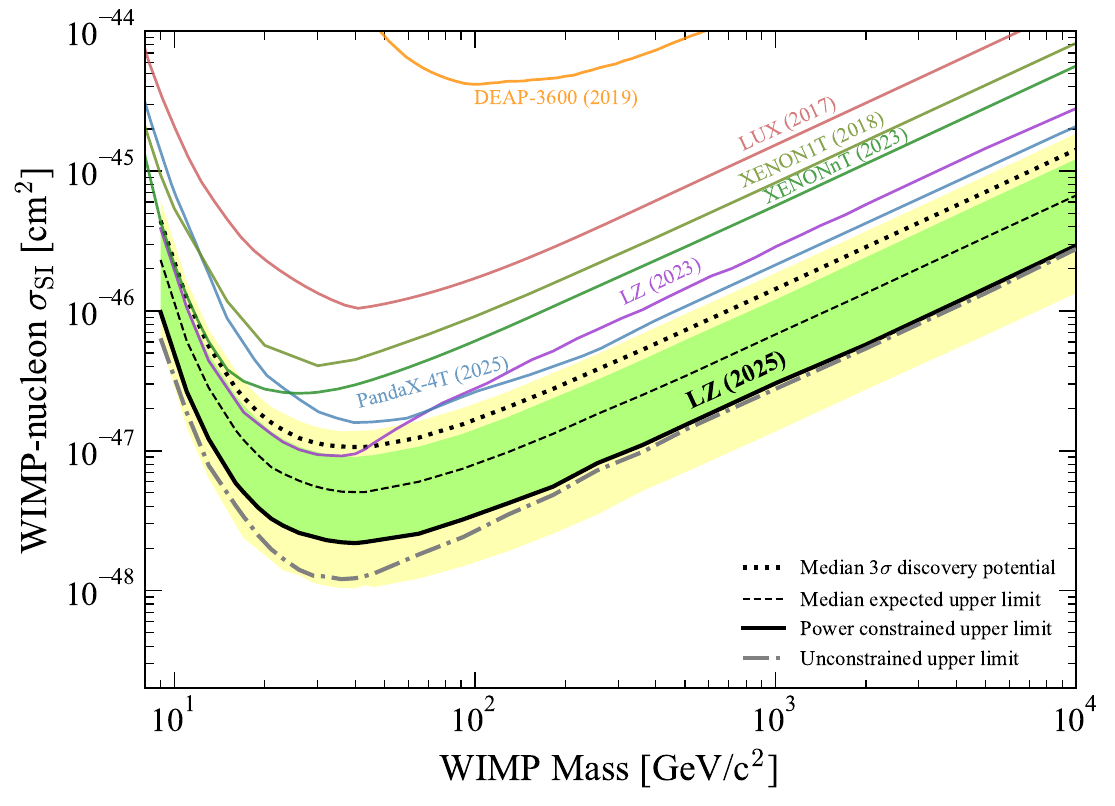}
\caption{
Upper limits (90\% C.L.) on the spin-independent WIMP-nucleon cross section as a function of WIMP mass from the combined WS2024+WS2022 analysis (280.0~live days) are shown with a solid black line, with a $-1\sigma$ power constraint applied. The gray dot-dash line shows the limits without the power constraint; green and yellow regions show the range of expected upper limits from 68\% and 95\% of background-only experiments, while the dashed black line indicates the median expectation, obtained with post-fit background estimates. The median $3\sigma$ observation potential from the post-fit model is shown as a dotted black line. Also shown are limits from WS2022 only~\cite{SR1WS}, PandaX-4T~\cite{PandaX:2024qfu}, LUX~\cite{akerib2017results}, all power constrained to $-1\sigma$; XENONnT~\cite{XENON:2023cxc}, reinterpreted with a $-1\sigma$ power constraint; XENON1T~\cite{collaboration2018dark}, and DEAP-3600~\cite{DEAP:2019yzn}.}
\label{fig:limit}
\end{figure}

Figure~\ref{fig:limit} shows the 90\% confidence level upper limit on the spin-independent WIMP-nucleon cross section as a function of mass following inference performed with a two-sided, unbinned profile likelihood ratio test statistic~\cite{cowan2011asymptotic}. The limit is power constrained at all WIMP masses to $1\sigma$ below the median expectation following the recommendation in Ref.~\cite{*[{}] [{; Here we employ a power constraint which uses the expected sensitivity of the experiment assuming no signal is present.}] baxter2021recommended}. The fluctuation, and therefore the size of the constraint, is largest between 20 and 50~GeV/$c^2$. This is a consequence of the inclusion of the WS2022 data from Ref.~\cite{SR1WS}, which exhibit an underfluctuation in this region. The limit for the WS2024 data alone can be found in the Supplemental Material~\cite{Supplemental}, with an approximate $-1\sigma$ fluctuation across the entire mass range, primarily attributed to an underfluctuation of the accidental coincidence background in the region of largest overlap with expected WIMP distributions. The minimum of the limit curve is $2.2 \times 10^{-48}$~cm$^2$ at 40~GeV/$c^2$ while that of the median expected limit is $5.1 \times 10^{-48}$~cm$^2$ at 40~GeV/c$^2$. The best median $3\sigma$ observation potential is $1.1\times10^{-47}$~cm$^2$ at 40~GeV/c$^2$. Results of searching for evidence of spin-dependent WIMP-nucleon couplings are discussed in the Appendix.

In summary, LZ has achieved limits on SI WIMP-nucleon interactions that surpass previous best exclusions by a factor of four or more for WIMP masses $>$9~GeV/$c^2$. A radon tag to target the dominant ER background from $^{214}$Pb decays has been demonstrated, with potential for optimization that will allow it to have greater impact on the sensitivities of future searches, particularly for physics with potential ER signals. The enhanced electron-ion recombination of $^{124}$Xe LL-shell DEC events was noted for the first time, and will be further investigated in future analyses. The experiment continues to take salted data towards a target 1000-day live time that will enable more sensitive searches for WIMP interactions and other new phenomena.

\emph{Acknowledgements} - The research supporting this work took place in part at the Sanford Underground Research Facility (SURF) in Lead, South Dakota. Funding for this work is supported by the U.S. Department of Energy, Office of Science, Office of High Energy Physics under Contract Numbers DE-AC02-05CH11231, DE-SC0020216, DE-SC0012704, DE-SC0010010, DE-AC02-07CH11359, DE-SC0015910, DE-SC0014223, DE-SC0010813, DE-SC0009999, DE-NA0003180, DE-SC0011702, DE-SC0010072, DE-SC0006605, DE-SC0008475, DE-SC0019193, DE-FG02-10ER46709, UW PRJ82AJ, DE-SC0013542, DE-AC02-76SF00515, DE-SC0018982, DE-SC0019066, DE-SC0015535, DE-SC0019319, DE-SC0024225, DE-SC0024114, DE-AC52-07NA27344, \& DE-SC0012447. This research was also supported by U.S. National Science Foundation (NSF); the UKRI’s Science \& Technology Facilities Council under award numbers ST/W000490/1, ST/W000482/1, ST/W000636/1, ST/W000466/1, ST/W000628/1, ST/W000555/1, ST/W000547/1, ST/W00058X/1, ST/X508263/1, ST/V506862/1, ST/X508561/1, ST/V507040/1, ST/W507787/1, ST/R003181/1, ST/R003181/2,  ST/W507957/1, ST/X005984/1, ST/X006050/1; Portuguese Foundation for Science and Technology (FCT) under award numbers PTDC/FIS-PAR/2831/2020; the Institute for Basic Science, Korea (budget number IBS-R016-D1); the Swiss National Science Foundation (SNSF)  under award number 10001549. This research was supported by the Australian Government through the Australian Research Council Centre of Excellence for Dark Matter Particle Physics under award number CE200100008. We acknowledge additional support from the UK Science \& Technology Facilities Council (STFC) for PhD studentships and the STFC Boulby Underground Laboratory in the U.K., the GridPP~\cite{faulkner2005gridpp,britton2009gridpp} and IRIS Collaborations, in particular at Imperial College London and additional support by the University College London (UCL) Cosmoparticle Initiative, and the University of Zurich. We acknowledge additional support from the Center for the Fundamental Physics of the Universe, Brown University. K.T. Lesko acknowledges the support of Brasenose College and Oxford University. The LZ Collaboration acknowledges the key contributions of Dr. Sidney Cahn, Yale University, in the production of calibration sources. This research used resources of the National Energy Research Scientific Computing Center, a DOE Office of Science User Facility supported by the Office of Science of the U.S. Department of Energy under Contract No. DE-AC02-05CH11231. We gratefully acknowledge support from GitLab through its GitLab for Education Program. The University of Edinburgh is a charitable body, registered in Scotland, with the registration number SC005336. The assistance of SURF and its personnel in providing physical access and general logistical and technical support is acknowledged. We acknowledge the South Dakota Governor's office, the South Dakota Community Foundation, the South Dakota State University Foundation, and the University of South Dakota Foundation for use of xenon. We also acknowledge the University of Alabama for providing xenon. For the purpose of open access, the authors have applied a Creative Commons Attribution (CC BY) license to any Author Accepted Manuscript version arising from this submission. Finally, we respectfully acknowledge that we are on the traditional land of Indigenous American peoples and honor their rich cultural heritage and enduring contributions. Their deep connection to this land and their resilience and wisdom continue to inspire and enrich our community. We commit to learning from and supporting their effort as original stewards of this land and to preserve their cultures and rights for a more inclusive and sustainable future.

\appendix
\section{Appendix - Spin Dependent Results}
\begin{figure}[htbp]
  \centering
  \includegraphics[ width=\linewidth]{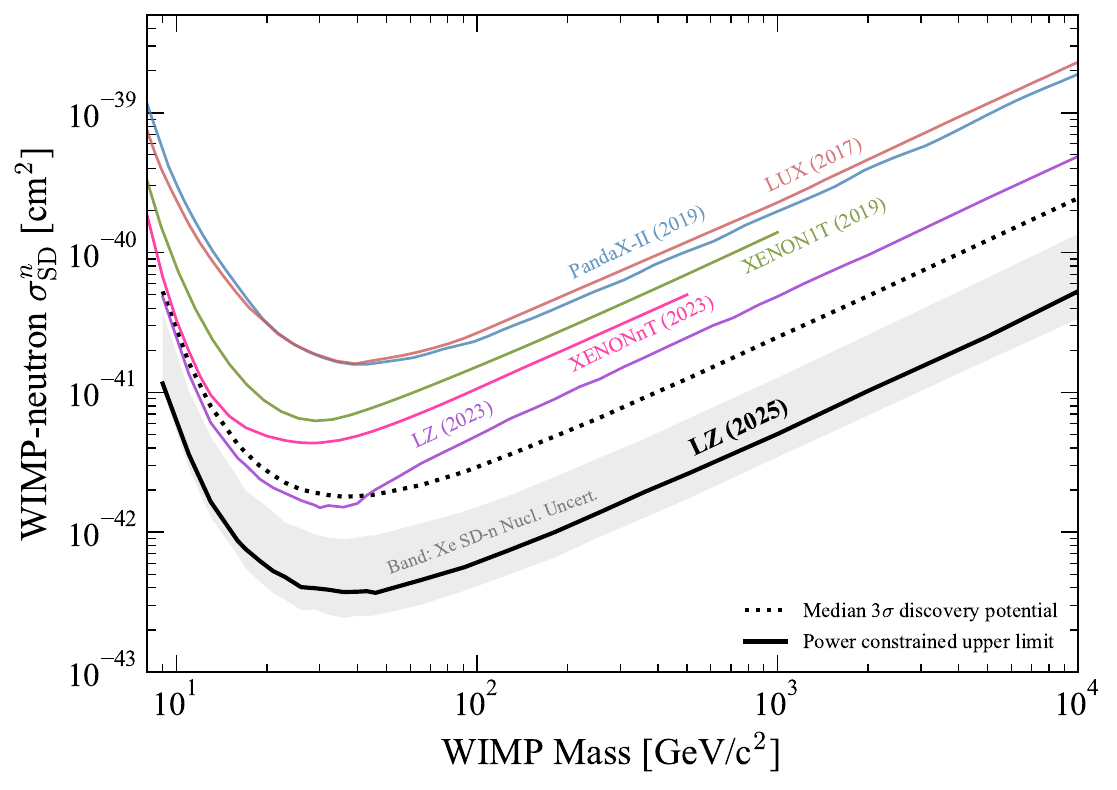}
  \caption{The 90\% confidence-level upper limit (solid-black curve) using the mean of the nuclear structure functions from~\cite{PhysRevD.102.074018}, allowing for like-to-like comparison with other results. 
  The gray band indicates the uncertainty due to nuclear modeling across models from~\cite{PhysRevD.102.074018,Pirinen:2019gap,PhysRevLett.128.072502} and applies similarly to the limit from all xenon-based experiments. The dotted line shows the median $3\sigma$ observation significance from the post-fit model.
  Also shown are the WS2022 only~\cite{SR1WS}, XENONnT~\cite{XENON:2023cxc} (reinterpreted with a $-1\sigma$ power constraint), XENON1T~\cite{collaboration2019constraining}, PandaX-II~\cite{xia2019pandax}, and LUX~\cite{akerib2017limits} limits.}
  \label{fig:limitplotSDn}
\end{figure}

\begin{figure}[htbp]
  \centering
  \includegraphics[ width=\linewidth]{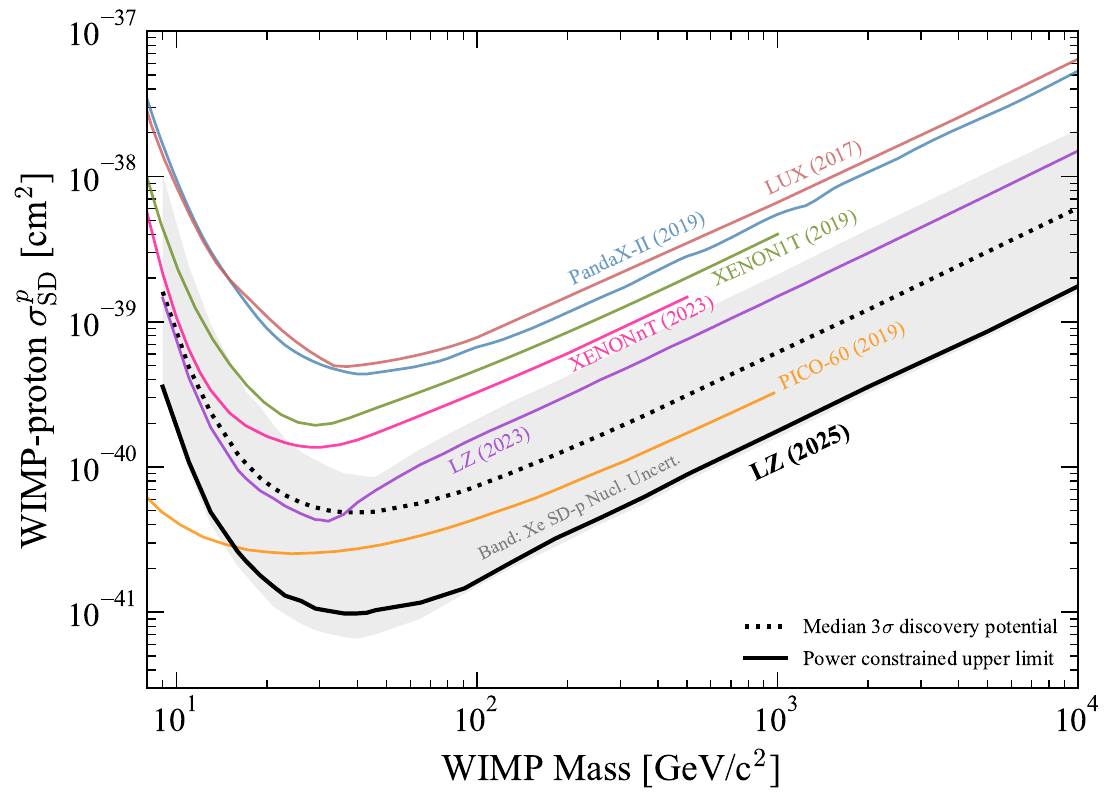}
  \caption{The 90\% confidence-level upper limit (solid-black curve) using the mean of the nuclear structure functions from~\cite{PhysRevD.102.074018}, allowing for like-to-like comparison with other results. 
  The gray band indicates the uncertainty due to nuclear modeling across models from~\cite{PhysRevD.102.074018,Pirinen:2019gap,PhysRevLett.128.072502} and applies similarly to the limit from all xenon-based experiments. 
  The dotted line shows the median $3\sigma$ observation significance from the post-fit model. 
  Also shown are the PICO-60~\cite{amole2019dark}, LZ WS2022 only~\cite{SR1WS}, XENONnT~\cite{XENON:2023cxc} (reinterpreted with a $-1\sigma$ power constraint), XENON1T~\cite{collaboration2019constraining}, PandaX-II~\cite{xia2019pandax}, and LUX~\cite{akerib2017limits} limits. The PICO-60 result relies on WIMP-scattering on the spin of the unpaired proton of $^{19}$F with minimal uncertainty.
  }
  \label{fig:limitplotSDp}
\end{figure}

In this Appendix we report limits on the spin-dependent WIMP-nucleon cross section using the same data selection, background modeling, and statistical inference techniques described in the main text. Results follow an identical procedure to that in Ref.~\cite{SR1WS}: signal models which describe scattering on \XeOneTwoNine\ (spin $1/2$, \SI{26.4}{\percent} natural abundance) and \XeOneThreeOne\ (spin $3/2$, \SI{21.2}{\percent} natural abundance)~\cite{XeAbundances} are constructed using the neutron- and proton-only nuclear structure functions and their uncertainties from Refs.~\cite{PhysRevD.102.074018,Pirinen:2019gap,PhysRevLett.128.072502}. The nominal limits use the mean structure functions from~\cite{PhysRevD.102.074018} and allow for a consistent comparison with previous limits from xenon-based experiments. As in Ref.~\cite{SR1WS}, an uncertainty band is calculated at each WIMP mass using the extrema of power-constrained limits obtained from the minimum and maximum nuclear structure function uncertainties provided in Refs.~\cite{PhysRevD.102.074018,Pirinen:2019gap,PhysRevLett.128.072502}.

The best-fit number of WIMP events is zero at all tested masses, from 9~GeV/$c^2$ to 100~TeV/$c^2$. The black line of Figure~\ref{fig:limitplotSDn} shows the 90\% confidence level nominal upper limit on the WIMP-neutron spin-dependent cross section as a function of mass, while the gray band shows the nuclear structure function uncertainty on the limit.  The minimum of the limit curve occurs at 46~GeV/$c^2$ at a cross section of $\sigma_{\mathrm{SD}}^{n} =$ \SI{3.7E-43}{cm\squared} (median expected limit at \SI{8.5E-43}{cm\squared} at 40~GeV/$c^2$), and a power constraint is applied at all tested masses. Figure~\ref{fig:limitplotSDp} shows the same for the WIMP-proton spin-dependent cross section. The minimum of the limit curve is at 40~GeV/$c^2$ at a cross section of $\sigma_{\mathrm{SD}}^p =$ \SI{9.8E-42}{cm\squared}, and a power constraint is applied between 9~GeV/$c^2$ and 91~GeV/$c^2$. The minimum of the median expected limit on $\sigma_{\mathrm{SD}}^p$ is \SI{2.3E-41}{cm\squared} and occurs at 40~GeV/$c^2$.

\bibliographystyle{apsrev4-2}
\bibliography{references}


\clearpage
\pagebreak

\widetext
\begin{center}
\textbf{\large Supplemental Materials}
\end{center}
\setcounter{equation}{0}
\setcounter{figure}{0}
\setcounter{table}{0}
\setcounter{page}{1}
\makeatletter
\renewcommand{\theequation}{S\arabic{equation}}
\renewcommand{\thefigure}{S\arabic{figure}}
\renewcommand{\thetable}{S\arabic{table}}

\section{Data Samples}
\begin{figure}[h]
    \centering
      \begin{tabular}{cc} \includegraphics[width=0.4\textwidth]{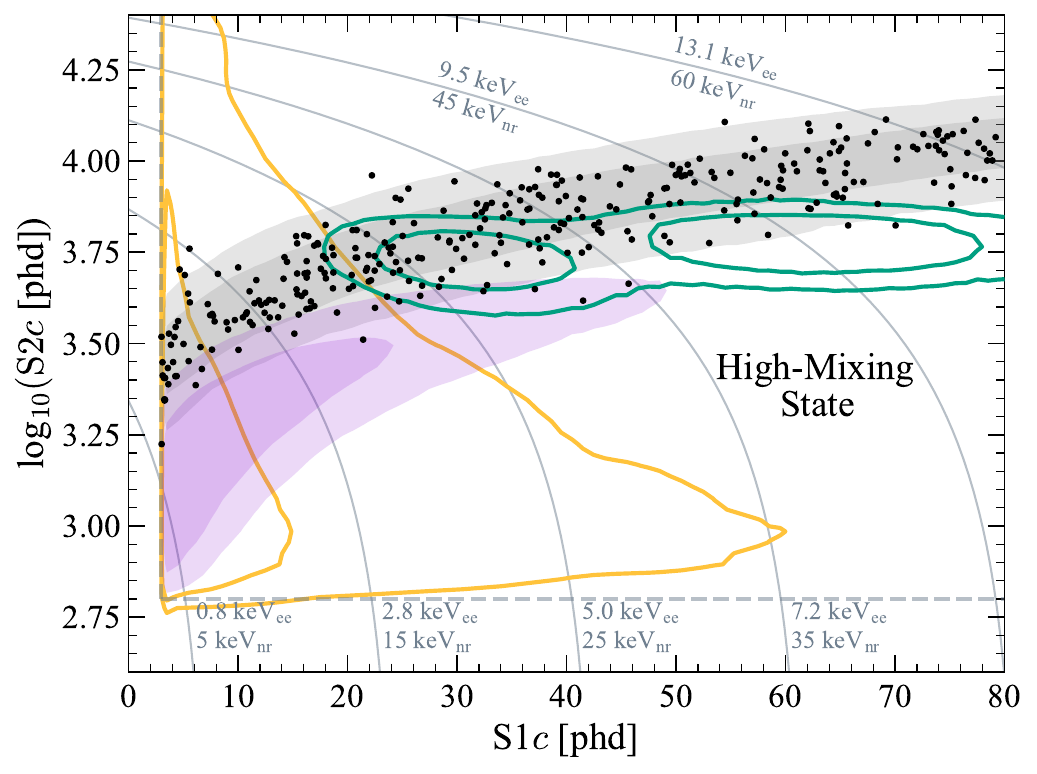} &  \includegraphics[width=0.4\textwidth]{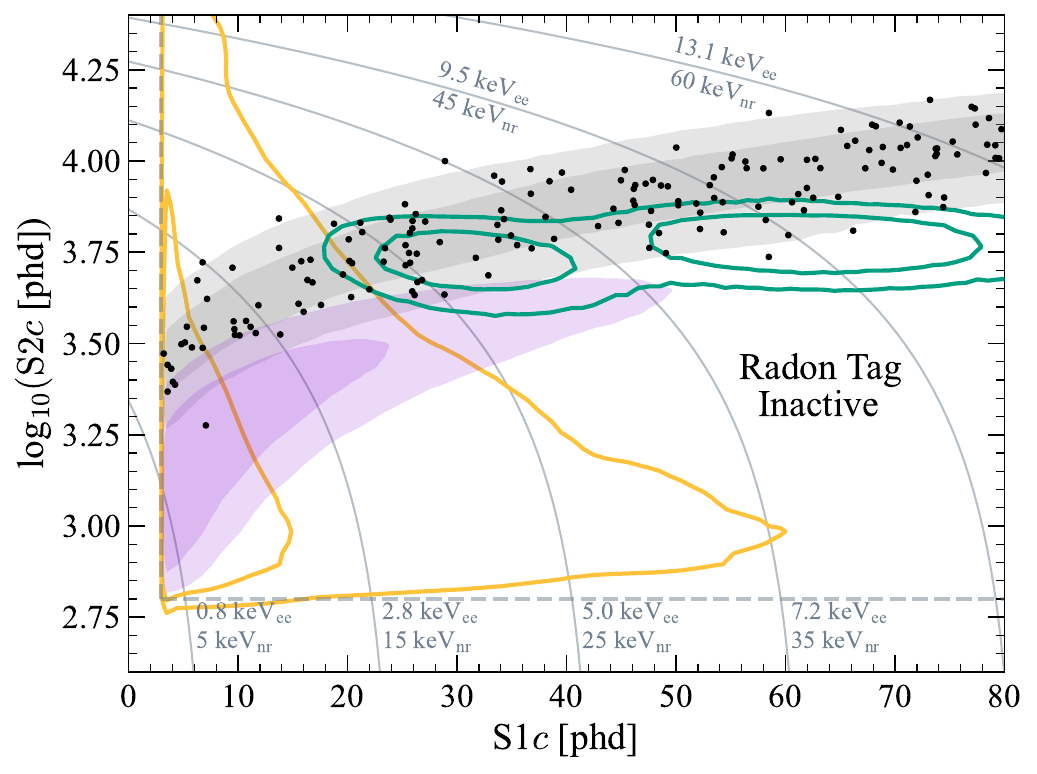} \\
    \includegraphics[width=0.4\textwidth]{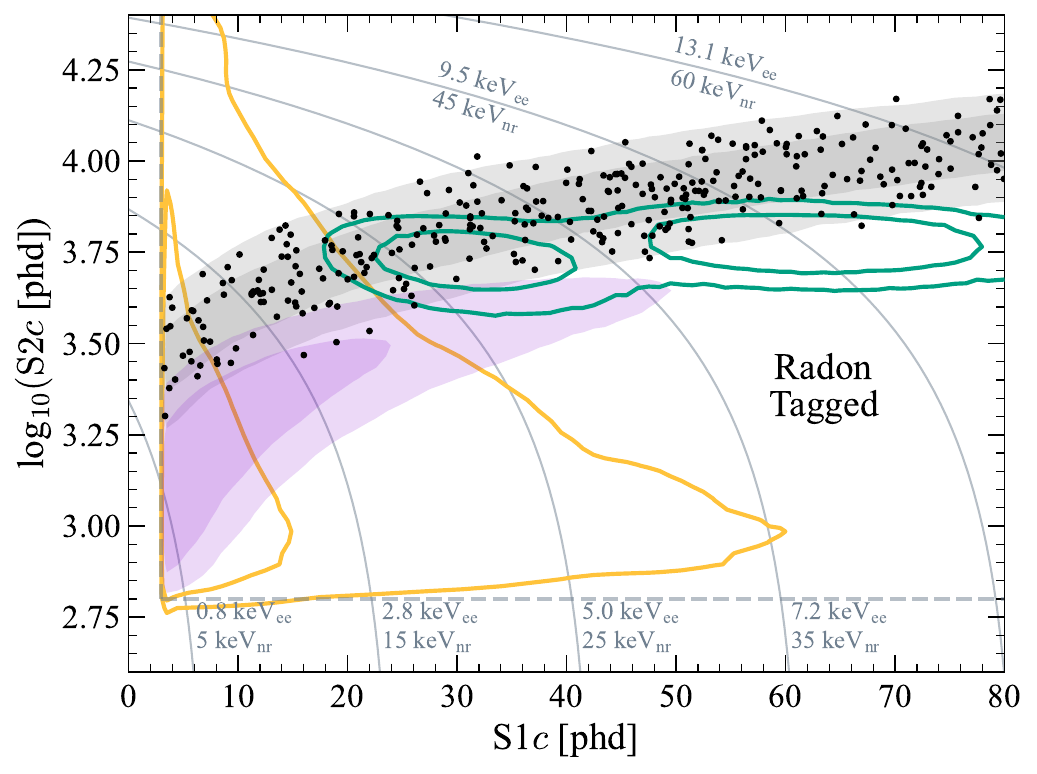} & \includegraphics[width=0.4\textwidth]{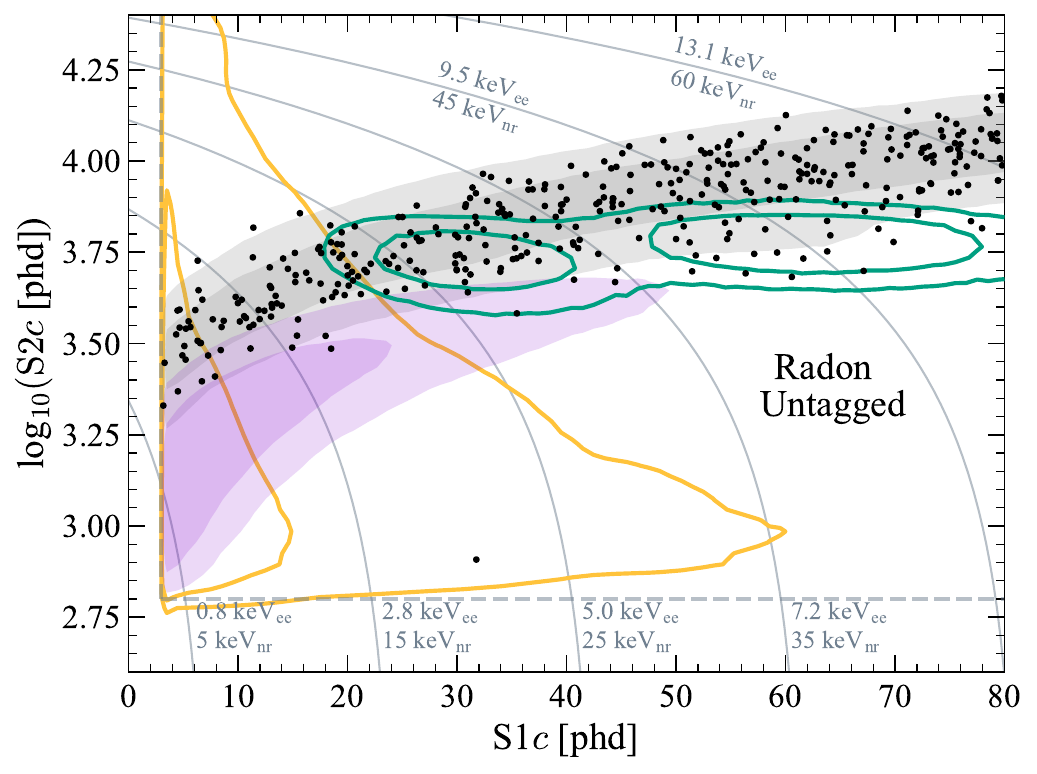} \\

    \includegraphics[width=0.4\textwidth]{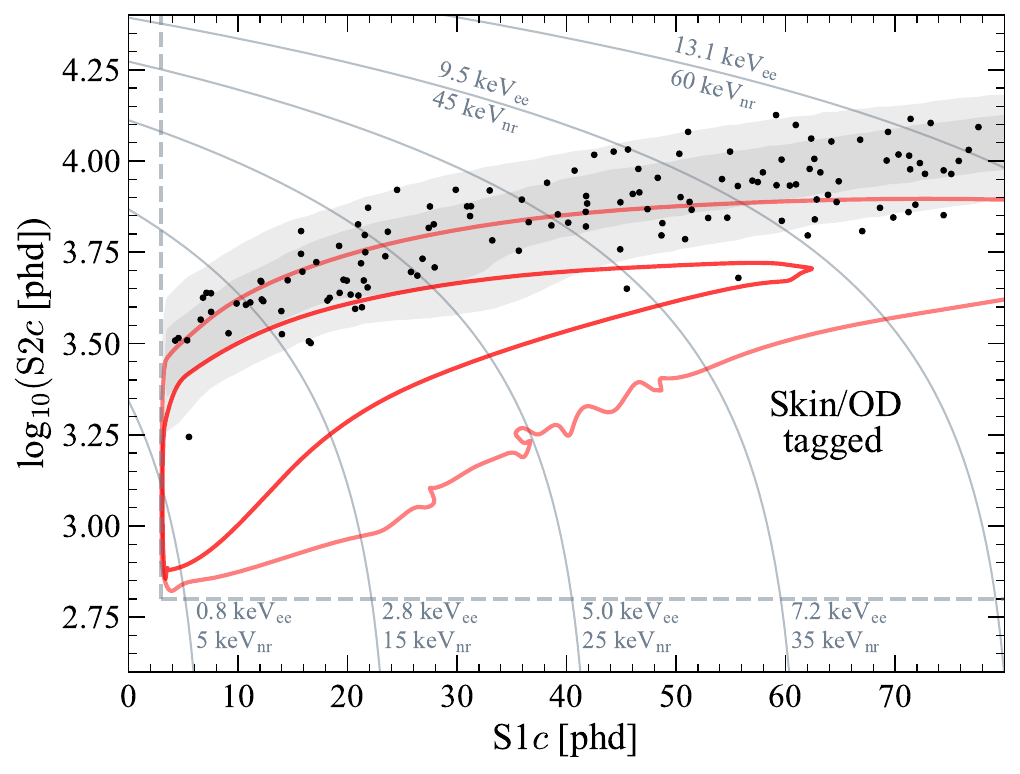}
 &    \includegraphics[width=0.4\textwidth]{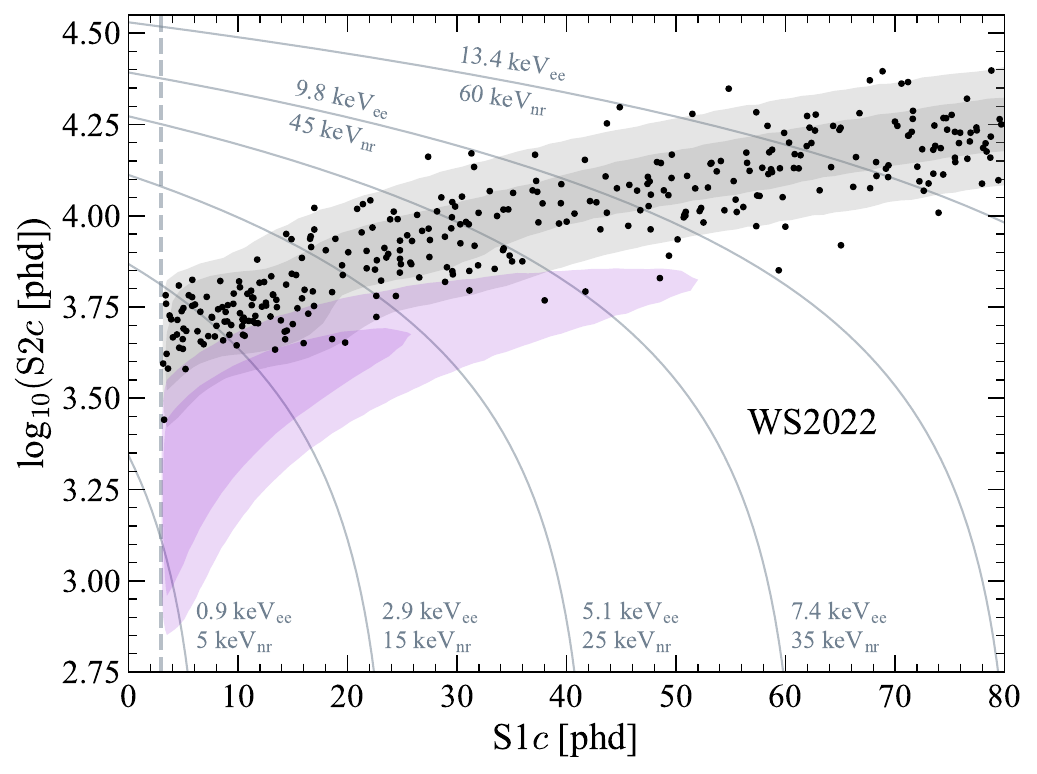}
    \end{tabular}
    \caption{
    Panels show the events used in the WS2024+WS2022 combined analysis organized into the various samples described in the main text. Model contours are the same as those in Fig.~\ref{fig:data}. Red contours in the veto tagged panel display the regions containing 68\% and 95\% of the distribution of tagged neutron backgrounds.}
    \label{fig:allsamples}
\end{figure}

Figure~\ref{fig:allsamples} shows the WS2024 and WS2022 events passing all data selection criteria grouped in their respective sample in the likelihood. Table~\ref{tab:exposures} lists the exposure for each sample in tonne-years.

\begin{table}[hbtp]
    \caption{Exposures of the six samples used in the likelihood given in tonne-years.}
    \label{tab:exposures}
    \centering
    \begin{tabular}{c c c c c c}
    \tabularnewline
    \hline
    \hline
    High-Mixing & Radon Tag & Radon & Radon & Veto & WS2022\\
    State & Inactive & Tagged & Untagged & Tagged & Only \tabularnewline
    \hline \tabularnewline[-2.2ex]
    0.6 & 0.6 & 0.3 & 1.8 & n/a & 0.9 \tabularnewline
    \hline
    \hline
    \end{tabular}
\end{table}

\section{Details of the Detector Response Model}
The LZ detector and xenon response models are implemented in a \textsc{nest}-based application that includes effects such as curved electron drift paths from field non-uniformities, finite position reconstruction resolution in the transverse $(x, y)$ and longitudinal $z$ directions, and position-dependence of the S1 and S2 signal sizes. The key detector parameters of the model are provided in Table~\ref{nest_det_params}. A header file for \textsc{nest}~v2.4.0 that reproduce the ER and NR response models used in this analysis is available online at \url{\supplementalurl}.

\begin{table}[h!]
    \centering
    \caption{Relevant \textsc{nest} detector parameters. The parameters in the top-half of the table are input parameters, while the bottom-half are results of \textsc{nest} calculations.\\}
    \begin{tabular}{l c c}
    \hline
    \hline
    Parameter & WS2024 & WS2022 \\ \hline
    $g_1$~[phd/photon]  &  0.112 & 0.114 \\
    $g_1^{\rm gas}$~[phd/photon] & 0.076 & 0.092 \\
    Effective GXe Extraction Field~[kV/cm] & 8.3 & 8.4 \\
    \hline
    Single Electron Size~[phd] & 46.9 & 58.5 \\
    Extraction Efficiency~[\%] & 72.5 & 80.5 \\
        $g_2$~[phd/electron] &~34.0 & 47.1 \\
    \hline
    \hline
    \end{tabular}
    \label{nest_det_params}
\end{table}

Note that NEST has multiple models for extraction efficiency versus gaseous xenon (GXe) electric field strength due to conflicting measurements in the literature~\cite{Xu:2019dqb}, and the default model provides the lower bound of the extraction efficiency for a given field strength. In order to reproduce the observed single electron size and measured extraction efficiency in LZ with this default model, a higher effective value for the GXe field magnitude is used, and thus the value reported in Table~\ref{nest_det_params} does not represent a physical field inside the LZ TPC. Extraction efficiency is measured as the ratio between the best-fit electroluminescence gain ($g_2$) value and the observed single electron signal size. 

In addition to the parameters below, the ER and NR mean yield and fluctuations models have been tuned to minimize the discrepancy between NEST and LZ calibration data. The ER mean yield parameters are provided in Table~\ref{nest_ER_params}, and the NR parameters are shown in Table~\ref{nest_NR_params}. Nomenclature for the individual parameters follows from~\cite{szydagis2022review}. There are detailed instructions for implementing these changes in the provided file referenced above.

\begin{table}[h!]
    \centering
    \caption{Default and best-fit model parameters for the \textsc{nest} ER mean charge yield model. The LZ values were found by fitting to the $^3$H and $^{14}$C band means. See Eq.~6 of~\cite{szydagis2022review} for more information regarding the individual $m_i$. \\}
    \begin{tabular}{l c c}
    \hline
    \hline
    Parameter & \textsc{nest} Default at 97~V/cm & LZ Value \\
    \hline
    $m_1$~[keV$^{-1}$] & 14.6 & 12.5 \\
    $m_2$~[keV$^{-1}$] & 77.3 & 85.0 \\
    $m_3$~[keV] & 0.8 & 0.6 \\
    $m_4$ & 2.1 & 2.1 \\
    $m_5$ [keV$^{-1}$] & 19.3 & 25.7 \\
    $m_6$ [keV$^{-1}$] & 0 & 0 \\
    $m_7$ [keV] & 78.0 &  59.7 \\
    $m_8$ & 4.3 & 3.7 \\
    $m_9$ & 0.33 & 0.28 \\
    $m_{10}$ & 0.08 & 0.11 \\
    \hline
    \hline
    \end{tabular}
    
    \label{nest_ER_params}
\end{table}

\begin{table}[h!]
    \centering
    \caption{Model parameters for the \textsc{nest} NR mean light and charge yields model. The LZ values were found by fitting to the calibrated DD band mean. See Eqs. (10), (12), and (13) of~\cite{szydagis2022review} for more information regarding the individual parameters. The last two parameters ($f_1$ and $f_2$) are not described in~\cite{szydagis2022review}, but are available as free parameters in \textsc{nest}, controlling the asymmetry of the sigmoidal energy dependence for the charge and light yields, respectively. See the \textsc{NEST} function \textsc{GetYieldNR(...)} for the usage of $f_1$ and $f_2$ in calculating NR yields~\cite{nest_2_4_0}. \\}
    \begin{tabular}{l c c}
    \hline
    \hline
    Parameter & \textsc{nest} Default & LZ Value \\
    \hline
    $\alpha$ [keV$^{1/\beta}$] & 11.0 & 10.2 \\
    $\beta$ & 1.1 & 1.1 \\
    $\gamma$ [(V/cm)$^{1/\delta}$] & 0.048 & 0.050 \\
    $\delta$ & -0.0533 & -0.0533 \\
    $\epsilon$ [keV] & 12.6 & 12.5 \\
    $\zeta$ [keV] & 0.30 & 0.29 \\
    $\eta$ & 2.0 & 1.9 \\
    $\theta$ [keV] & 0.30 & 0.32 \\
    $\iota$ & 2.0 & 2.1 \\
    $p$ & 0.50 & 0.51 \\
    \hline
    $f_1$ & 1.0 & 0.996 \\ 
    $f_2$ & 1.0 & 0.999 \\
    \hline
    \hline
    \end{tabular}
    
    \label{nest_NR_params}
\end{table}

\section{WS2024-Only Limit}
\begin{figure}[htbp]
  \centering
  \includegraphics[ width=0.5\linewidth]{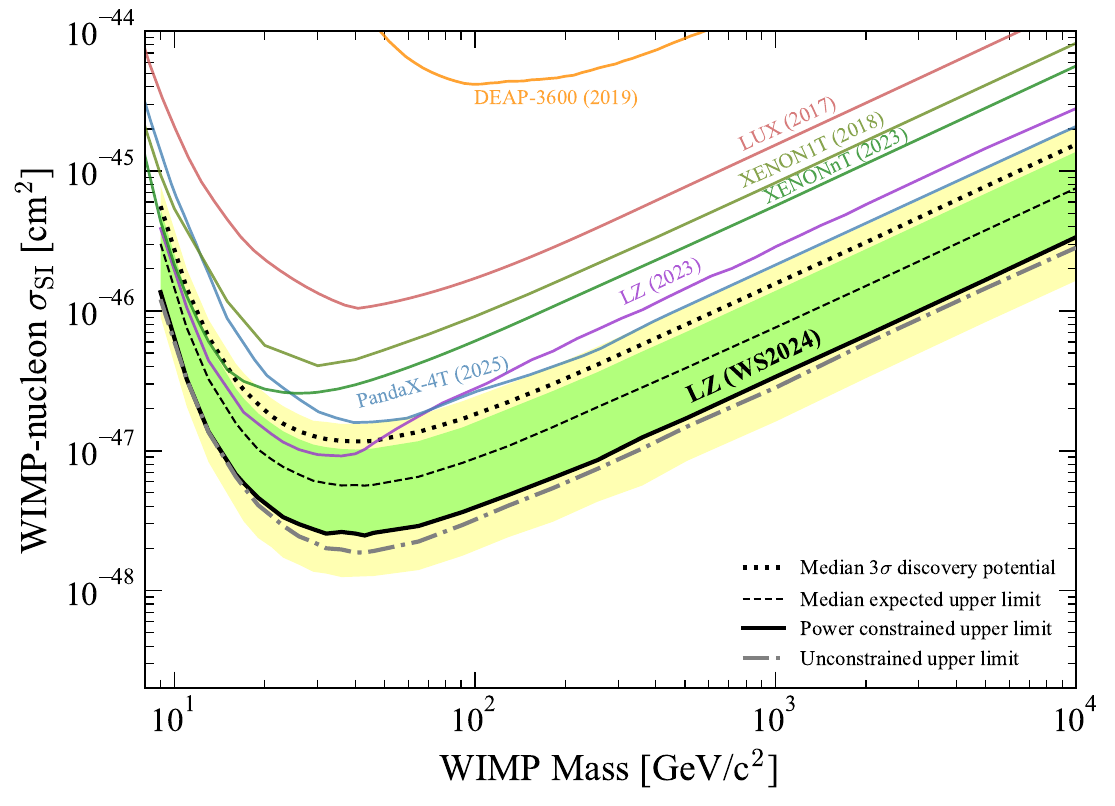}
  \caption{Upper limits (90\% C.L.) on the spin-independent WIMP-nucleon cross-section as a function of WIMP mass from WS2024 (220.0~live days) only are shown with a solid black line, with a $-1\sigma$ power constraint applied. The gray dot-dash line shows the limits without the power constraint; green and yellow regions show the range of expected upper limits from 68\% and 95\% of background-only experiments, while the dashed black line indicates the median expectation, obtained with post-fit background estimates. The median $3\sigma$ observation significance from the post-fit model is shown as a dotted black line. Also shown are limits from WS2022 only~\cite{SR1WS}, PandaX-4T~\cite{PandaX:2024qfu}, LUX~\cite{akerib2017results}, all power constrained to $-1\sigma$; XENONnT~\cite{XENON:2023cxc}, reinterpreted with a $-1\sigma$ power constraint; XENON1T~\cite{collaboration2018dark}, and DEAP-3600~\cite{DEAP:2019yzn}.}
  \label{fig:ws2024only}
\end{figure}

Figure~\ref{fig:ws2024only} shows the SI WIMP-nucleon cross section upper limit obtained using the data and models pertaining to the WS2024, 3.3~tonne-year exposure only.

\pagebreak
\section{Limits in Terms of Number of WIMP Events}

\begin{figure}[htbp]
  \centering
  \begin{minipage}{0.49\textwidth}
        \centering
        \includegraphics[width=0.98\textwidth]{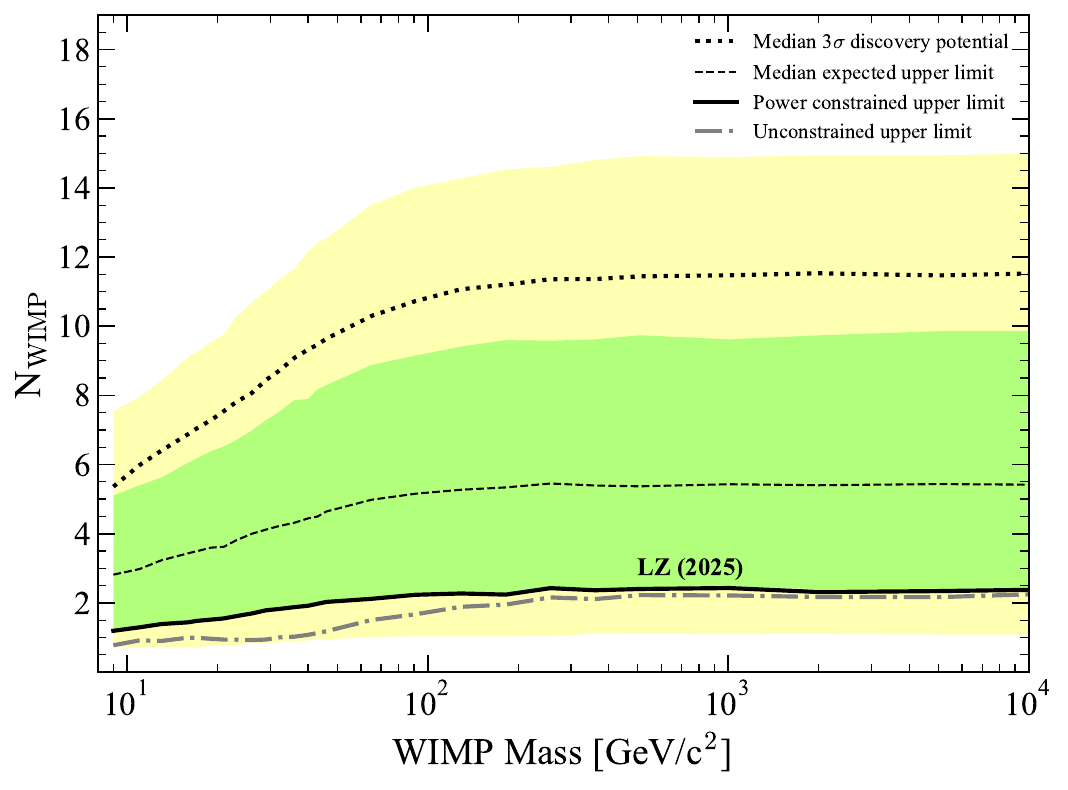}
  \end{minipage}\hfill
  \begin{minipage}{0.49\textwidth}
        \centering
        \includegraphics[width=0.98\textwidth]{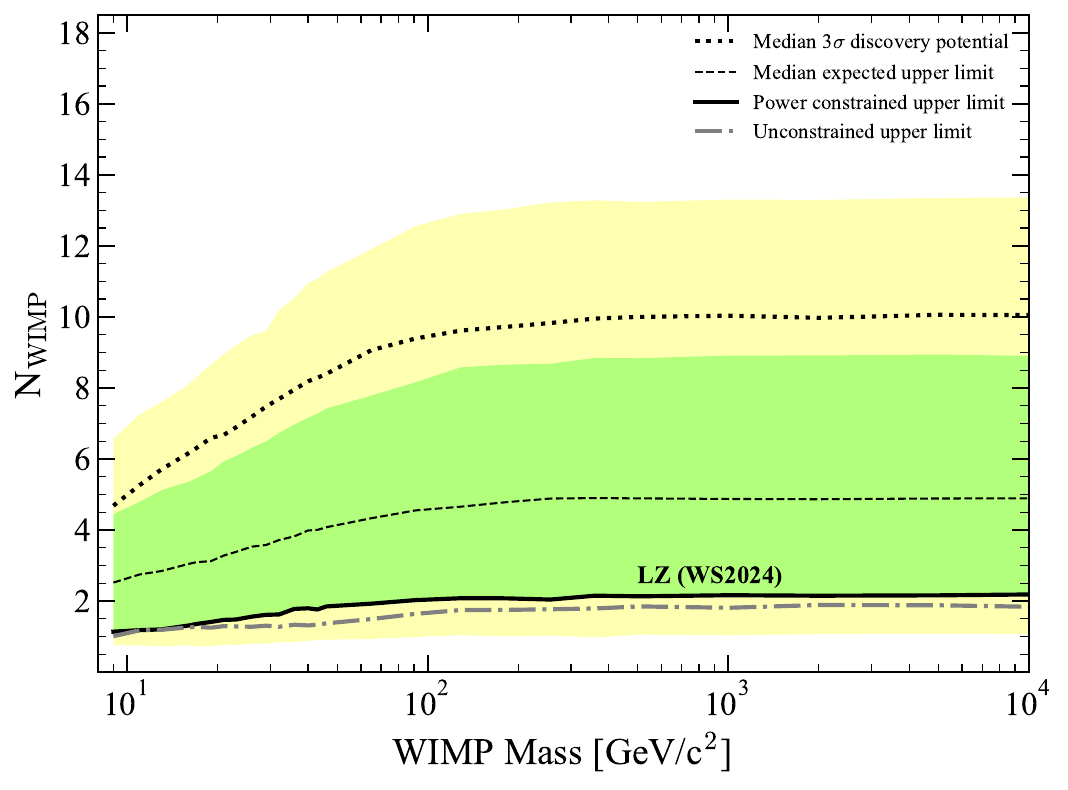}
  \end{minipage}\hfill
  \caption{The \SI{90}{\percent} confidence limit (solid line) for the number of WIMP events versus WIMP mass in the spin-independent case for the combined analysis (left) and WS2024 only (right). The dot-dash line shows the limit before applying the power constraint described in the text. The dashed line shows the median sensitivity and the dotted line shows the median $3\sigma$ discovery sensitivity. The green and yellow bands are the $1\sigma$ and $2\sigma$ sensitivity bands. }
  \label{fig:nwimp}
\end{figure}

Figure~\ref{fig:nwimp} presents the WS2022+WS2024 combined and WS2024-only upper limits in number of events. 

\vspace{-0.03cm}
\RevA{\section{Parameters of the Salting Model}
As discussed in the main text, the salt is drawn from a distribution consisting of the sum of a WIMP-like exponential and a flat spectrum, as shown in Fig.~\ref{fig:saltdist}. The parameter $E_{max}$ sets the maximum energy of the flat distribution and is drawn from a Gaussian distribution with mean $\mu_{max} = 250$~keV and sigma $\sigma_{max} = 25$~keV. The parameter $E_\mu$ is chosen to mimic the recoil spectrum induced by a WIMP with mass $M_\chi$, which is in turn drawn from a Gaussian with mean $\mu_\chi$ = 80~GeV/c$^2$ and width $\sigma_\chi$ = 20~GeV/c$^2$. The normalization is drawn from a uniform distribution that spans from zero up to a rate that would be consistent with the upper limits from Ref.~\cite{SR1WS}. 

\begin{figure}[htbp]
  \centering
  \includegraphics[trim={0cm 7cm 0 5cm},clip, width=0.45\linewidth]{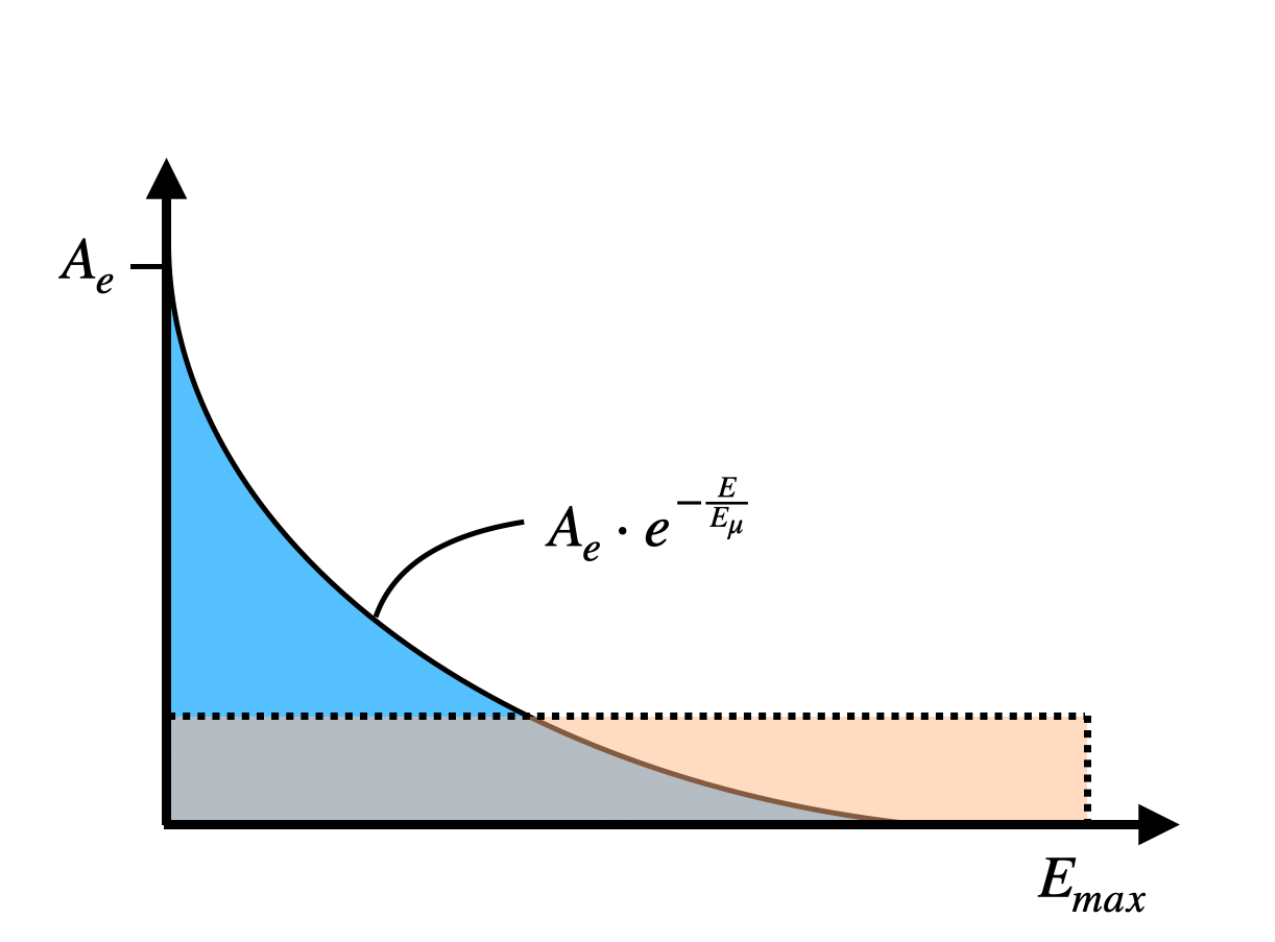}
  \caption{Distribution from which the salt events are drawn. The parameters are described in the text.}
  \label{fig:saltdist}
\end{figure}

The actual normalization, exponential, and flat spectrum parameters used are still hidden to the collaboration as unveiling them would harm bias mitigation of ongoing analyses. The allowed ranges were chosen to ensure that salt events can plausibly span the WIMP-search energy range, and also reach the highest recoil energies that might be used in a future higher-energy analysis, such as for effective field theories of dark matter.}

\end{document}